\documentclass[a4paper,final]{appolb}
\usepackage[utf8]{inputenc}
\usepackage[usenames,dvipsnames]{color}
\usepackage[dvips]{graphicx}

\usepackage[bookmarks]{hyperref}
        \hypersetup{colorlinks,  linktocpage, citecolor=MidnightBlue, filecolor=Blue, linkcolor=MidnightBlue, urlcolor=MidnightBlue}
\usepackage[numbers, sort&compress]{natbib} %pour avoir une génération automatique de réf. bib [1-6] au lieu de lister [1, 2, 3, 4, 5, 6].
\usepackage{hypernat} 
\newcommand {\pT}       {\ensuremath{p_{\mathrm{t}}}}
\newcommand {\meanpT}   {\ensuremath{\langle p_{\mathrm{t}} \kern-0.1em\rangle}} 
\newcommand {\sqrtS}    {\ensuremath{\sqrt{s}}}
\newcommand {\sqrtSnn}  {\ensuremath{\sqrt{s_{NN}}}}

\newcommand {\rap}      {\mbox{$\left | y \right | $}}

\newcommand {\cTau}     {\ensuremath{c.\tau}}

\newcommand {\dNdy}     {\ensuremath{\mathrm{d}N/\mathrm{d}y}}

\newcommand {\dNdptdy}  {\ensuremath{\mathrm{d^{2}}N/\mathrm{d}\pT\mathrm{d}y }}

\newcommand {\pp}       {\mbox{$\mathrm {p\kern-0.05em p}$}}

\newcommand {\AuAu}     {\ensuremath{\mbox{Au--Au} }}

\renewcommand {\AA}     {\ensuremath{\mbox{A--A}}}

\newcommand {\mass}     {\mbox{\rm MeV$\kern-0.15em /\kern-0.12em c^2$}}
\newcommand {\tev}      {\mbox{${\rm TeV}$}}

\newcommand {\mom}      {\mbox{\rm GeV$\kern-0.15em /\kern-0.12em c$}}
\newcommand {\gmom}     {\mbox{\rm GeV$\kern-0.15em /\kern-0.12em c$}}
\newcommand {\mmass}    {\mbox{\rm MeV$\kern-0.15em /\kern-0.12em c^2$}}
\newcommand {\gmass}    {\mbox{\rm GeV$\kern-0.15em /\kern-0.12em c^2$}}
\newcommand {\mmom}     {\mbox{\rm MeV$\kern-0.15em /\kern-0.12em c$}}

\newcommand {\cm}       {\mbox{${\rm cm}$}}

\newcommand {\dg}       {\mbox{$\kern+0.1em ^\circ$}}

\newcommand{\piMinus}           {\mbox{$\mathrm {\pi^-}$}}

\newcommand{\Kminus}    {\mbox{$\mathrm {K^-}$}}

\newcommand{\proton}    {\mbox{$\mathrm {p}$}}

\newcommand{\rmLambdaZ}         {\mbox{$\mathrm {\Lambda}$}}

\newcommand{\rmLambda}          {\mbox{$\mathrm {\Lambda}$}}
\newcommand{\rmAlambda}         {\mbox{$\mathrm {\overline{\Lambda}}$}}

\newcommand{\rmXi}      {\mbox{$\mathrm {\Xi}$}}
\newcommand{\rmXiM}     {\mbox{$\mathrm {\Xi^{-}}$}}
\newcommand{\rmXiPM}    {\mbox{$\mathrm {\Xi^{\pm}}$}}
\newcommand{\rmAxiP}    {\mbox{$\mathrm {\overline{\Xi}^{+}}$}}
\newcommand{\rmXis}     {\mbox{$\mathrm {\Xi^{-}+\overline{\Xi}^{+}}$}}

\newcommand{\rmOmega}   {\mbox{$\mathrm {\Omega}$}}
\newcommand{\rmOmegaM}  {\mbox{$\mathrm {\Omega^{-}}$}}
\newcommand{\rmOmegaPM} {\mbox{$\mathrm {\Omega^{\pm}}$}}
\newcommand{\rmAomegaP} {\mbox{$\mathrm {\overline{\Omega}^{+}}$}}
\newcommand{\rmOmegas}  {\mbox{$\mathrm {\Omega^{-}+\overline{\Omega}^{+}}$}}

\newcommand{\Jpsi}      {\mbox{J\kern-0.05em /\kern-0.05em$\psi$}}

\begin{document}
\eqsec  % uncomment this line to get equations numbered by (sec.num)

% numérotation des lignes
% \modulolinenumbers[2]
%\pagewiselinenumbers
% \switchlinenumbers   % permet de mettre les num. sur les marges extérieures
% \linenumbers

\title{Production of multi-strange baryons in 7 TeV proton-proton collisions with ALICE%
\thanks{Presented at Strangeness In Quark Matter 2011, 18-24 Sept. 2011, Krakow.}%
}
\author{Antonin MAIRE\thanks{Now at Physikalisches Institut, Universit\"at Heidelberg, Germany.}, for the ALICE Collaboration
\address{Institut Pluridisciplinaire Hubert Curien, Strasbourg, France}
}

%\author{,  \\ {\footnotesize }.}

\maketitle
\begin{abstract}

In the perspective of comparisons between proton-proton and heavy-ion physics, understanding the production mechanisms (soft and hard) in \pp{} that lead to strange particles is of importance. 
Measurements of charged multi-strange (anti-)baryons (\rmOmegaPM{} and \rmXiPM) are presented for \pp{} collisions at \sqrtS{} = 7 \tev. This report is based on results obtained by ALICE (A Large Ion Collider Experiment) from the 2010 data-taking.\\
\indent Taking advantage of the characteristic cascade-decay topology, the identification of \rmXiM, \rmAxiP, \rmOmegaM{} and \rmAomegaP can be performed, over a wide range of momenta (e.g. from 0.6 to 8.5 \gmom{} for \rmXiM, with the present statistics analysed). 
The production at central rapidity ($\rap < 0.5$) as a function of transverse momentum, \dNdptdy, is presented. These results are compared to PYTHIA Perugia 2011 predictions.

\end{abstract}

%%%---------------------------------------------------------------------------------------------------------------
%
\section{Introduction}
\label{sec:Intro}
%--------

In heavy-ion (\AA) as well as in proton-proton (\pp) collisions, the measurements related to strange hadrons constitute unique tools to study the physics of the strong interaction.
In this respect, given their strangeness content, the charged multi-strange baryons -- \rmXiM($dss$), \rmAxiP($\bar{d}\bar{s}\bar{s}$), \rmOmegaM($sss$), \rmAomegaP($\bar{s}\bar{s}\bar{s}$) -- are of certain importance. 
The interest in strangeness and specifically in \rmXi{} and \rmOmega{} can be explained mainly by two reasons :
\begin{itemize}
        \item the initial system formed by the projectiles is free from strange valence quarks, thus the strange quarks that compose the strange hadrons of the final state have to be produced during the process of the collisions;
        \item due to the identification via weak-decay topology reconstruction, the multi-strange baryons can be studied over a large momentum range, typically from \pT~$\approx 0.5~\mom$ up to $\approx 10~\mom$. The resulting \pT{} spectra then cover the region dominated by the soft processes and reach the energy scale where hard scattering mechanisms may prevail.
\end{itemize}

A significant part of the investigation in \AA{} physics depends on our understanding of the production mechanisms (soft and hard) in the \pp{} system, meaning as they occur in the benchmark system.
The measurement of multi-strange particle spectra in \pp{} collisions is the topic of this report.

%The ALICE experiment~\cite{Alessandro:2006yt} is well-suited for such spectrum measurements, due to a low \pT~cut-off and excellent particle identification (PID) capabilities. The low \pT~cut-off is made possible by the low magnetic field applied in the central barrel ($\leq 0.5~T$) and the low material budget in this mid-rapidity region ($13\%$ of radiation length~\cite{Hippolyte:2009xz}). The PID capabilities are supplied by a set of detectors utilizing diverse techniques (energy loss, time of flight, transition radiation, Cerenkov effect).

%%%---------------------------------------------------------------------------------------------------------------
%
\section{Data analysis}
\label{sec:DataAnalysis}

%----------------------------------------------
\subsection{Data collection and detector setup}
\label{ssec:DataColl-DetSetUp}
%--------

The data presented here are from a minimum bias (MB) sample of \pp{} collisions at \sqrtS{} = 7 \tev, 
collected at the LHC \cite{evans2008lhcJINST} with the ALICE experiment \cite{alice2008pstationExpmtInJINST} during summer 2010. 
The entire statistics presently analysed stands for about $165 \times 10^{6}$ MB interactions.

The study makes use of the two ALICE main tracking detectors placed at mid-rapidity, covering the full azimuth : 
the Inner Tracking System, composed of 6 cylindrical layers of high-resolution silicon detectors~\cite{alice2010AlignmentITSinJINST}, 
and the cylindrical Time Projection Chamber (TPC)~\cite{alice2010TPCInNIM}.
%The whole is embedded in the large L3 solenoidal magnet which provides a nominal magnetic field of $0.5~T$. 

\subsection{Topological reconstruction and signal extraction}
\label{ssec:TopoReco-SigExtr}
%--------

The multi-strange hadron identification is performed using a combination of displaced-vertex finding, invariant mass analysis and particle identification (PID) method at the single track level (in this analysis, compatibility selection based on the energy loss measurements from the TPC).

The reconstruction of the \rmXiM, \rmAxiP, \rmOmegaM, \rmAomegaP{} particles hinges on their respective charged weak decays, the so-called \emph{cascade} structures.
For each species of interest, the main characteristics and utilized decay channels are listed in Table \ref{tab:Maire-CascDecayChannel}. The anti-baryons are similarly reconstructed via the decay channel involving the charge conjugates. 

 \begin{table}[!hbt]
         \begin{center}
                 \begin{tabular}{lccccc}
                \noalign{\smallskip}\hline \noalign{\smallskip}
                        Particles   &   mass ($\mmass$)   &  \cTau{} (\cm)  & charged decay channel & B.R. \\
                \noalign{\smallskip}\hline \noalign{\smallskip}
                        $\rmLambdaZ$ ($uds$)              & $1115.7$   & $7.89$ &  $\rmLambdaZ \rightarrow \proton + \piMinus$             & 63.9\% \\
                        %$\rmAlambdaZ$ ($\overline{uds}$)  & $1115.68$   & $7.89$ &  $\rmAlambdaZ \rightarrow \pbar + \piPlus$ & 63.9\% \\
                %\noalign{\smallskip}\hline \noalign{\smallskip}
                        $\rmXiM$  ($dss$)                 & $1321.7$   & $4.91$ &  $\rmXiM \rightarrow  \rmLambdaZ+ \piMinus $         & 99.9\% \\
                        %$\rmAxiP$~($\overline{dss}$)      & $1321.71$   & $4.91$ &  $\rmAxiP \rightarrow \rmAlambdaZ+ \piPlus $       & 99.9\% \\
                %\noalign{\smallskip}\hline \noalign{\smallskip}
                        $\rmOmegaM$ ($sss$)               & $1672.5$   & $2.46$ &  $\rmOmegaM \rightarrow  \rmLambdaZ+ \Kminus$   & 67.8\% \\
                        %$\rmAomegaP$ ($\overline{sss}$)   & $1672.45$  & $2.46$ &  $\rmAomegaP \rightarrow  \rmAlambdaZ+ \Kplus$ & 67.8\% \\
                \noalign{\smallskip}\hline \noalign{\smallskip}
                 \end{tabular}
        \caption{Main characteristics of the reconstructed particles~\cite{ParticleDataGroup2010}.}
        \label{tab:Maire-CascDecayChannel}
         \end{center}
 \end{table}

The guidelines of the reconstruction algorithm, consisting of pairing a \rmLambda~or \rmAlambda~baryon with an additional particle (\emph{V0} structure combined with a so-called \emph{bachelor} track), are sketched in \cite{maire2010multistrangeInPpHQ} and detailed in \cite{alice2010strange900GeV}.
The protocol to extract the signal counts per \pT{} interval (bin) is also presented in \cite{alice2010strange900GeV}.

%%%---------------------------------------------------------------------------------------------------------------
%
\section{Results}
\label{sec:Section-Results}

%%%---------------------------------
\subsection{Corrected \pT{} spectra}
\label{ssec:CorrectedSpectra}
%--------

Due to the large statistics available for the 7 \tev~\pp~data sample, \rmXiM{} and \rmAxiP{} but also \rmOmegaM{} and \rmAomegaP{} can be studied distinctively. 
At central rapidity (\mbox{$\rap < 0.5$}), the overall raw signal amounts to about $105 \times 10^3$ counts for \rmXiM, as well as for \rmAxiP, and about $5 \times 10^3$ counts for \rmOmegaM{} as well as for its anti-particle.

\begin{figure}[!hbt]
               \centering
                        \includegraphics[scale=0.5]{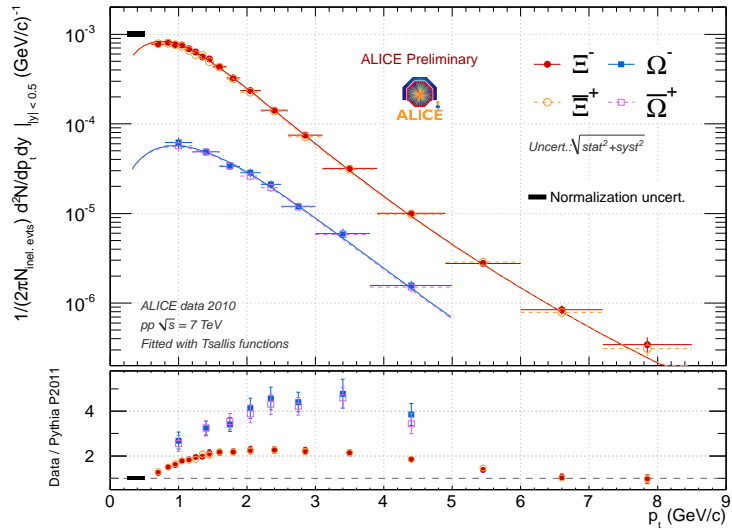}
                        \caption{Corrected spectra \dNdptdy{} as a function of transverse momentum, for \rmXiM, \rmAxiP, \rmOmegaM, \rmAomegaP{} measured at central rapidity (\rap $< 0.5$) in \pp{} inelastic events at \sqrtS{} = 7 \tev. The upper panel shows the corrected spectra for the four different species together with their fits by a Tsallis function. The lower panel presents the ratio between the measured data and the predictions returned by the tune Perugia 2011 (P2011) of PYTHIA.}
                        \label{fig:Maire-Fig1-CorrectedSpectra}  
        \end{figure}

The upper part of the Figure \ref{fig:Maire-Fig1-CorrectedSpectra} shows the four corrected spectra.
They are normalised to the number of inelastic events (INEL).
Both \rmXi{} spectra range from \pT{} = 0.6 \gmom{} to 8.5 \gmom, while the \rmOmega{} spectra are measured between 0.8 \gmom{} and 5.0 \gmom.
In order to extract the integrated yields, the data points are fitted with a Tsallis function \cite{alice2010strange900GeV}. 
The fits result in a good description of data ($\chi^2/NDF$ close to unity) and are further used to extrapolate the spectra in the non-measured low \pT{} region ($\approx 22 \%$ of the total \dNdy{} for \rmXiM{} or \rmAxiP, $\approx 26 \%$ for \rmOmegaM{} or \rmAomegaP).

Note that, provided the difference in the normalisation scheme chosen by ALICE (INEL) and CMS (Non-Single Diffractive, NSD) \cite{cms2011strangeParticlesInPP900GeVand7TeV}\footnotemark, the (\rmXis) spectra by both experiments are found to be in agreement within the limit of uncertainties.

\footnotetext{The NSD cross-section stands for about 80 \% of the total INEL one \cite{poghosyan2011diffractiveProcessInPP7TeVQM2011}.}

%%%-------------------------------------------
\subsection{Comparison to PYTHIA Perugia 2011}
\label{ssec:ComparisonPYTHIA}
%--------

A significant part of our understanding of \pp{} collisions is based on description by the Monte Carlo models (MC). This is especially the case in the \emph{soft} regime (\pT $<$  2-3 \gmom).
The confrontation with experimental measurements is necessary, as it spurs further improvements of such phenomenological approaches.

The lower part of the Figure \ref{fig:Maire-Fig1-CorrectedSpectra} shows the ratio between data and MC predictions.
The data are here compared to the tune Perugia 2011 (tune S350 \cite{skands2010perugiaTunes}) of the PYTHIA model \cite{Sjostrand2006pythia64}.
Although this specific tune provides an improved description of data with respect to earlier tunes (tunes Z2 and Perugia 0) \cite{chinellato2011multistrangeInALICEQM2011}, it can still be seen that MC underestimates the measured spectra, up to a factor $\sim$ 2 for the charged \rmXi, $\sim$ 5 for the \rmOmega{}.
However it should be noted for \rmXiM{} (\rmAxiP) that there is an agreement between MC and data at \pT{} $>$ 6 \gmom, where the fragmentation regime may already be reached.

\begin{figure}[!htb]
               \centering
                        \includegraphics[scale=0.46]{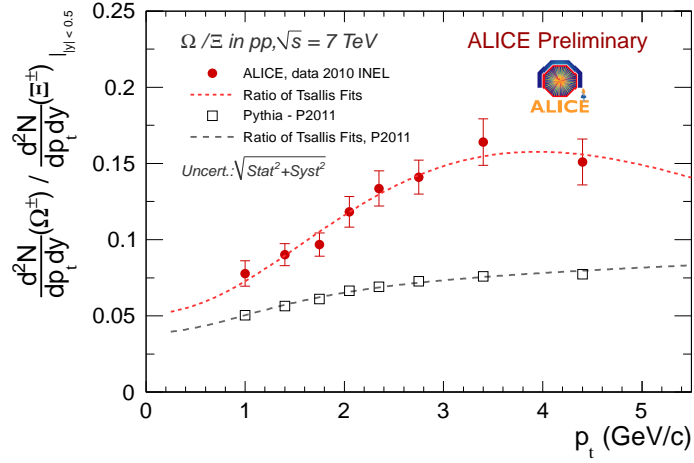}
                        \caption{Ratio of \dNdptdy{} between (\rmOmegaM+\rmAomegaP) and (\rmXiM+\rmAxiP) as a function of transverse momentum, for \pp{} collisions at \sqrtS{} = 7 \tev. The ratio is given for ALICE data as well as PYTHIA Perugia 2011 (P2011). As a consistency check, the ratio of the Tsallis functions fitted to the data or to the model predictions is also shown.}
                        \label{fig:Maire-Fig2-RatioOmegaOverXi-Pt}  
        \end{figure}

The Figure \ref{fig:Maire-Fig2-RatioOmegaOverXi-Pt} provides another viewpoint on the comparison of data to MC. It computes the ratio of (\rmOmegas) to (\rmXis) as a function of \pT. As before, it can be noted the model tends to underpredict the particle ratio, by a factor $\sim$2.
Regarding the data ratio itself, it turns out the ratio first rises with \pT, before seemingly saturating at a value of 0.15, which could suggest that the hierarchy between \rmOmega{} and \rmXi{} production becomes constant at high \pT.

        \begin{figure}[!hbt]
               \centering
                        \includegraphics[scale=0.46]{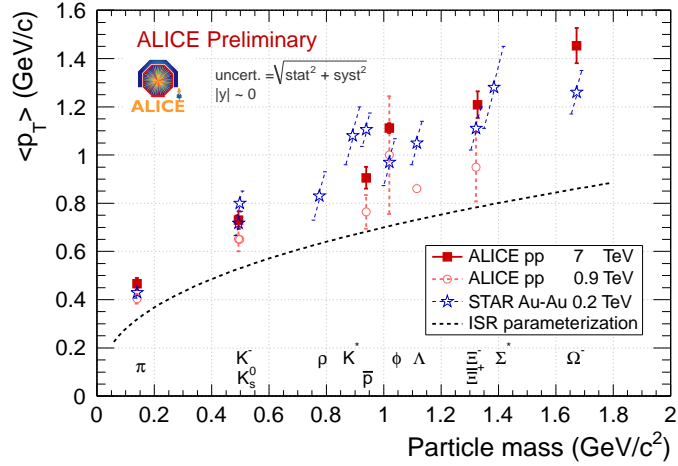}
                        \caption{Mean \pT{} values (\meanpT) for different species identified by the ALICE experiment in \pp{} collisions, at \sqrtS{} = 0.9 \cite{alice2011idSpectraPP900GeV, alice2010strange900GeV} and 7 \tev. These \pp{} values are compared to the ISR parametrisation (performed from the $\pi$, K, p species in \pp{} collisions at \mbox{\sqrtS{} =  0.025 \tev}) and STAR measurements in central \AuAu{} collisions at \mbox{\sqrtSnn{} = 0.2 \tev{}} \cite{star2007strangeInPP}.}
                        \label{fig:Maire-Fig3-ALICEmeanPt-PPcollisions}  
        \end{figure}

%%%---------------------------------------------
\subsection{Mean \pT{} for identified particles}
\label{ssec:MeanPt}
%--------

The Figure \ref{fig:Maire-Fig3-ALICEmeanPt-PPcollisions} presents the ALICE preliminary values obtained for the mean \pT{} (\meanpT) of various identified particles, as measured in \pp{} collisions at 7 \tev. For comparison, the data are plotted together with results published by ALICE for \pp{} data at \sqrtS{} = 0.9 \tev{} \cite{alice2011idSpectraPP900GeV, alice2010strange900GeV} as well as by STAR for the central \AuAu{} collisions at \sqrtSnn{} = 0.2 \tev{} \cite{star2007strangeInPP}. For a given particle studied in \pp{} collisions, it appears that the \meanpT{} rises with \sqrtS. At \mbox{\sqrtS{} = 7 \tev}, the \pp{} data come to reach the values obtained in the most central \AA{} collisions at RHIC, raising the question of the physical similarities between these different systems at different energies.

%%%---------------------------------------------------------------------------------------------------------------
%
\section{Summary}
\label{sec:Ccl}
%--------

With the results presented in this report, the ALICE experiment provides the first LHC measurement of \rmOmegaM{} and \rmAomegaP{} at central rapidity. 
Together with \rmXiM{} and \rmAxiP{} results, it actually enables an extension of the excitation function in a \pp{} system.
This is of importance to define the \pp{} baseline necessary for the studies related to the so-called \emph{strangeness enhancement} \cite{nicassio2011XiandOmegaInPbPb276TeV}. Besides, although the tune Perugia 2011 is a priori the current most suitable PYTHIA tune for the description of hyperons in \pp{} collisions at LHC energies, it is observed that it underestimates the data at intermediate \pT.

% Some measurements have already been performed at previous and current facilities. These include both \ppbar~collisions (S$\mathrm{p\overline{p}}$S, Tevatron) and \pp~collisions (RHIC), with centre of mass energies $\sqrt{s}$ ranging from 0.2~\tev~up to 1.96 \tev~\cite{star2007strangeInPP,Ansorge:1989ba,CDFrunII:2010h}. The LHC, in operation since November 2009, can extend the existent 0.9-\tev~measurements made by the UA5 collaborations in \ppbar, and perform new measurements at $\sqrt{s}$ = 7~\tev, beyond the Tevatron energies.

%______________________________________________________________________________
%________________________________________________________________References

% \bibliographystyle{MyBibStyle3}
% Style perso défini pour la bibliographie de la thèse (Oct, 2009)
% \bibliographystyle{alpha}
% \bibliographystyle{cparalleless}
% \bibliographystyle{wmaainf}  % style bibliographique de la thèse de Jeff, je crois = a chercher
\bibliographystyle{unsrtnat}

\end{document}